\documentclass[journal,comsoc]{IEEEtran}
\usepackage[T1]{fontenc}

\usepackage[dvipsnames]{xcolor}

\usepackage{cite}
\usepackage{graphicx}
\graphicspath{{Figures/}}
\usepackage{subcaption}
\usepackage{amsmath}
\usepackage[cmintegrals]{newtxmath}
\usepackage{amsfonts}
\usepackage{gensymb}
\usepackage{bm}

\usepackage[ruled,vlined,resetcount,linesnumbered]{algorithm2e}

\usepackage{array}

\usepackage[bottom]{footmisc}

\begin{document}
%
\title{\huge{Tensor-Decomposition-based Hybrid Beamforming Design for mmWave OFDM Massive MIMO Communications}}

%
%

\author{\vspace{-0.2cm}
	Guilherme~Martignago~Zilli,~\IEEEmembership{Student Member,~IEEE,}
        Wei-Ping~Zhu,~\IEEEmembership{Senior Member,~IEEE}
\thanks{G. M. Zilli and W.-P. Zhu. are with the Department of Electrical and Computer Engineering, Concordia University, Montreal, QC, Canada, H3G 1M8. E-mail: guilherme.m.zilli(at)gmail.com; weiping(at)ece.concordia.ca.}%
\thanks{This work is partly supported by the Fonds de Recherche du Qu\'ebec - Nature et Technologies (FRQNT) and the Natural Sciences and Engineering Research Council of Canada (NSERC).}
\thanks{This work has been submitted to the IEEE for possible publication. Copyright may be transferred without notice, after which this version may no longer be accessible.}}

\maketitle

\begin{abstract}
	In this paper, we propose a novel joint hybrid precoder and combiner design for maximizing the average achievable sum-rate of single-user orthogonal frequency division multiplexing millimeter wave massive MIMO systems.
	We formulate the analog precoder and combiner design as a constrained Tucker2 decomposition and solve it by using the projected alternate least square method. Such a formulation allows maximizing the sum of the effective baseband gains over all subcarrier, while suppressing the interference among the data streams in the same subcarrier. In turn, the digital precoder and combiner are obtained from the effective baseband channel's singular value decomposition on a per-subcarrier basis. Numerical simulation results show that the proposed method outperforms other existing designs.
	\vspace{-0.2cm}
\end{abstract}


%
\IEEEpeerreviewmaketitle


\vspace{-0.4cm}
\section{Introduction}
\label{sec:intro}

	\IEEEPARstart{H}{ybrid} beamforming (HBF) and massive MIMO are key enabling technologies for 5G and beyond systems to explore the large bandwidths available in millimeter wave (mmWave) and terahertz (THz) bands and provide higher~data rates and lower-latency~\cite{Xiao2017,Giordani2020}. HBF combines a high-dimensional analog beamforming and a low-dimensional digital beamforming, thus providing an appealing trade-off by achieving near-optimal performance, supported by the high array and multiplexing gains of massive MIMO, while preserving the reduced complexity and power consumption due to the lower number of RF chains~\cite{Pi2011,Rappaport2013,Heath2016}. 
	However, the coupling between analog and digital beamforming and the constant-modulus constraint imposed by the analog beamforming hardware renders non-convex nonlinear designing problems.
	
	The promising results obtained by initial works on narrowband mmWave massive MIMO HBF design~\cite{Ayach2014,Sohrabi2016,Zilli2020}, along with the large bandwidths available in mmWave and THz frequencies encouraged researchers to explore frequency-selective broadband channels. In particular, they focused on the
	orthogonal frequency-division multiplexing (OFDM), where the frequency-selective channel is decomposed into a set of non-interfering parallel frequency-flat narrowband sub-channels, allowing precoders~and combiners to be designed for each narrowband channel, i.e., each subcarrier~\cite{Bolckei2002}.
	Nonetheless, OFDM HBF design is challenging since the analog beamforming is shared among all subcarriers~\cite{Yu2016,Sohrabi2017,Tsai2019,Chiang2018}. 
	
	In~\cite{Yu2016}, the authors described an alternating optimization approach to minimize the sum of the distances between the HBF and optimal unconstrained beamforming of each subcarrier. Their approach, however, requires \textit{a priori} computation of the optimal unconstrained beamforming.
	Under the assumption that the channel covariance matrices at different subcarriers are asymptotically equal and share nearly the same set of eigenvectors, the authors in~\cite{Sohrabi2017} designed the analog beamforming as a narrowband analog beamforming, using the average channel covariance matrices over all subcarriers, while the digital precoder was obtained on a per-subcarrier basis aiming to maximize the overall spectral efficiency.
	In~\cite{Tsai2019}, the authors designed the analog precoder and combiner by extracting the phases of the eigenvectors of the averaged channel covariance matrix and the averaged conjugate-transposed-channel covariance matrix, respectively.
	Finally, in~\cite{Chiang2018}, the authors presented a beam selection approach, where the analog beamforming vectors are selected from orthogonal codebooks.
	
	Given the multidimensional structure of OFDM channels, tensor decomposition emerged as a promising tool to handle OFDM system design, particularly for channel estimation~\cite{Araujo2019,Ruble2020}, and digital precoder design~\cite{Almeida2008}. Tensor decomposition has also been considered in HBF design for narrowband multiuser systems~\cite{Liu2017}. However, to the best of your knowledge, it is yet to be explored in the OFDM HBF design.
		
	Here, we propose a novel two-stage HBF design method for single-user OFDM mmWave massive MIMO communication systems, where we first design the analog beamforming, and then its digital counterpart. Specifically, by using a tensor representation of the OFDM channel (i.e., by stacking together the channel matrices of all subcarriers into a tensor), we formulate the analog beamforming design problem as a constrained Tucker2 tensor decomposition and solve it by the projected alternate least square (ALS) method. Such a formulation allows us to maximize the sum of the effective baseband gains over all subcarriers, while suppressing the interference among data streams within the same subcarrier.
	The optimal digital beamforming is obtained from the effective baseband channel' singular value decomposition (SVD) on a per-subcarrier basis.
	Finally, we provide extensive simulation results, which have shown that the proposed design achieves near-optimal sum-rates, outperforming others HBF designs in the literature.
		 
\vspace{-0.2cm}	 
\section{System Model and Problem Formulation}
\label{sec:system_model}	
				
\begin{figure}[!t]
			\centering
			\hspace{-0.3cm}
			\includegraphics[width=1.02\columnwidth]{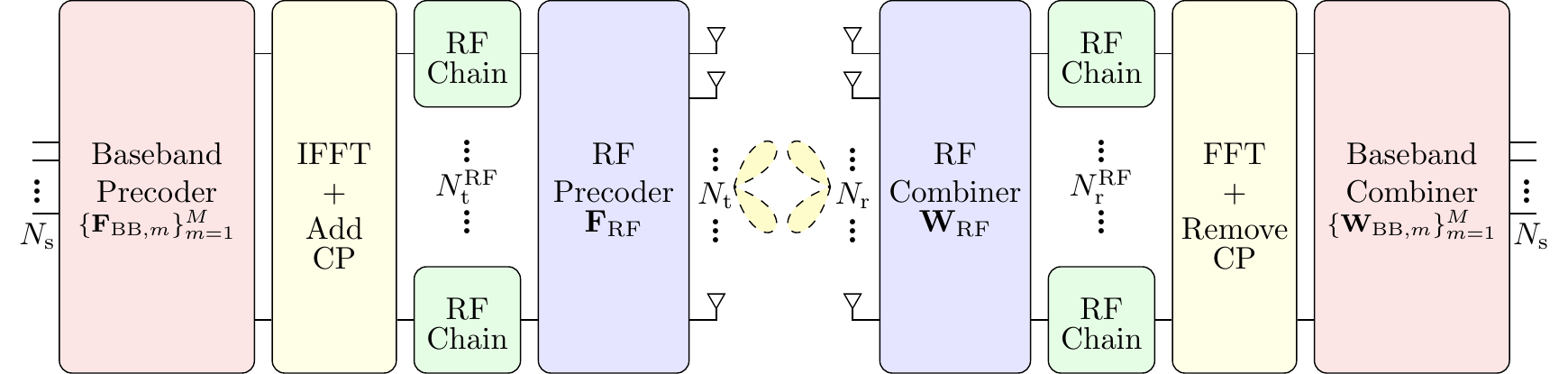}
			\caption{\small{System diagram. }}
			\vspace{-0.5cm}
			\label{fig:system_model}
		\end{figure}

		We consider that the transmitter, having $N_{\mathrm{t}}$ antenna elements and $N_{\mathrm{t}}^{\mathrm{RF}}$ RF chains, and the receiver, having $N_{\mathrm{r}}$ antenna elements and $N_{\mathrm{r}}^{\mathrm{RF}}$ RF-chains, use $M$ subcarriers to communicate through $N_{\mathrm{s}}$ data streams per subcarrier, as depicted in Fig.~\ref{fig:system_model}.
		We assume the transmitted symbol vector for the $m^{\mathrm{th}}$ subcarrier $\mathbf{s}_{m} \in \mathbb{C}^{N_\mathrm{s} \times  1}$ has i.i.d components and unity power, i.e., $\mathbb{E} [\mathbf{s}_{m} \mathbf{s}_{m}^{\mathrm{H}}] = \frac{1}{N_\mathrm{s}} \mathbf{I}_{N_\mathrm{s}}$.
		At the transmitter, the symbol vector of each subcarrier is digitally precoded on a per-subcarrier basis, using the $m^{\mathrm{th}}$ subcarrier digital precoder $\mathbf{F}_{\mathrm{BB},m} \in \mathbb{C}^{N_\mathrm{t}^{\mathrm{RF}}~\times~N_\mathrm{s}}$, and goes through the OFDM modulator, where the cyclic prefix addition and the inverse fast Fourier transform operations take place~\cite{Bolckei2002}.
		The OFDM symbols are precoded on the analog domain through a single analog precoder $\mathbf{F}_{\mathrm{RF}} \in \mathbb{C}^{N_\mathrm{t}~\times~N_\mathrm{t}^{\mathrm{RF}}}$, common to all subcarriers. The analog precoder is implemented using phase shifters, and thus imposes a constant modulus constraint (i.e., $| \mathbf{F}_{\mathrm{RF}}(m,n) | = 1/\sqrt{N_\mathrm{t}}~\forall m,n$).
		At the receiver, the received signal is combined on the analog domain by the analog combiner matrix $\mathbf{W}_{\mathrm{RF}} \in \mathbb{C}^{N_\mathrm{r}~\times~N_\mathrm{r}^{\mathrm{RF}}}$ (with  $| \mathbf{W}_{\mathrm{RF}}(m,n) | = 1/\sqrt{N_\mathrm{r}}~\forall m,n$), common to all subcarriers, and goes through the OFDM demodulation. Finally, the demodulated signals are combined by the digital combiner matrix $\mathbf{W}_{\mathrm{BB},m} \in \mathbb{C}^{N_\mathrm{t}^{\mathrm{RF}}~\times~N_\mathrm{s}}$ on a per-subcarrier basis.
		Thus, considering $\mathbf{H}_{m}$ the frequency domain channel matrix of the $m^{\mathrm{th}}$ subcarrier, the received symbol corresponding to the $m^{\mathrm{th}}$ subcarrier is
		\vspace{-0.1cm}
		\begin{equation}
			\label{eq:OFDM_receivedSignal_combiner}
			\mathbf{\hat{s}}_{m} = \sqrt{\rho} \mathbf{W}_{\mathrm{BB},m}^{\mathrm{H}}  \mathbf{W}_{\mathrm{RF}}^{\mathrm{H}} \mathbf{H}_{m} \mathbf{F}_{\mathrm{RF}} \mathbf{F}_{\mathrm{BB},m} \mathbf{s}_{m} + \mathbf{W}_{\mathrm{BB},m}^{\mathrm{H}} \mathbf{W}_{\mathrm{RF}}^{\mathrm{H}} \mathbf{n}
			\vspace{-0.1cm}
		\end{equation}
		where $\rho$ is the average  power of the received signal and  $\mathbf{n} \in \mathbb{C}^{N_\mathrm{r} \times 1}$ is the noise vector, such that $\mathbf{n}$ is i.i.d. $\mathcal{CN}(\mathbf{0},\sigma_\mathrm{n}^2 \mathbf{I}_{N_\mathrm{r}})$.
		
		In practical deployments, linear detection techniques are preferred due to the high computational complexity of nonlinear techniques. 
		In this case, the $N_\mathrm{s}$ data streams are treated independently at the receiver and thus, the sum-rate of all data streams is a reasonable performance metric.
		The achievable sum-rate of the $m^\mathrm{th}$ subcarrier is
		\vspace{-0.15cm}
		\begin{equation}
			\label{eq:OFDM_achievedSumRate}
			R_{\mathrm{sum},m} = \sum_{k=1}^{N_\mathrm{s}} \log_2 \left( 1 + \gamma_{k,m} \right)
			\vspace{-0.15cm}
		\end{equation}
		where $\gamma_{k,m}$ is the SINR of the $k^\mathrm{th}$ data stream in the $m^\mathrm{th}$ subcarrier, defined as
		\vspace{-0.1cm}
		\begin{equation}
		\label{eq:OFDM_streamSINR}
		\hspace{-0.02cm}
		\gamma_{k,m} = \frac{\frac{\rho}{N_\mathrm{s}} | \mathbf{W}_{m}(:,k)^{\mathrm{H}} \mathbf{H}_{m} \mathbf{F}_{m}(:,k)|^2 }{\frac{\rho}{N_\mathrm{s}} \sum_{ i=1, i\neq k }^{N_\mathrm{s}} |\mathbf{W}_{m}(:,k)^{\mathrm{H}} \mathbf{H}_{m} \mathbf{F}(:,i) |^2 + \sigma_\mathrm{n}^2 \| \mathbf{W}_{m}(:,k) \|^2}
		\hspace{-0.3cm}
		\vspace{-0.0cm}
		\end{equation}
		with $\mathbf{F}_{m} = \mathbf{F}_{\mathrm{RF}} \mathbf{F}_{\mathrm{BB},m}$ and $\mathbf{W}_{m} = \mathbf{W}_{\mathrm{RF}} \mathbf{W}_{\mathrm{BB},m}$ being the effective hybrid precoder and combiner for the $m^\mathrm{th}$ subcarrier.

		We adopt the extended Saleh-Valenzuela channel model, which corresponds to the sum of the contributions of $N_{\textrm{cl}}$ scattering clusters with $N_{\textrm{ray}}$ propagation paths per cluster.
		For OFDM systems, the frequency-selective channel is separated into $M$ narrowband frequency-flat sub-channels with non-interfering subcarriers. The frequency domain channel matrix for the $m^{\mathrm{th}}$ subcarrier is given by~\cite{Lee2014}
		\vspace{-0.1cm}
		\begin{equation}
			\label{eq:OFDM_ChannelModel}
			\mathbf{H}_{m} = \sqrt{\frac{N_\mathrm{r} N_\mathrm{t}}{N_{\mathrm{cl}}  N_{\mathrm{ray}}}} \sum_{i=0}^{N_{\mathrm{cl}}-1} \sum_{l=0}^{N_{\mathrm{ray}}-1} \alpha_{il} \mathbf{a}_\mathrm{r} (\phi_{il}^\mathrm{r},\theta_{il}^\mathrm{r}) \mathbf{a}_\mathrm{t}^{\mathrm{H}} (\phi_{il}^\mathrm{t},\theta_{il}^\mathrm{t}) e^{\frac{-j 2 \pi i (m-1)}{M}}
			\vspace{-0.0cm}
		\end{equation}	
		where $\alpha_{il}\thicksim \mathcal{CN}(0, \sigma_{\alpha,i}^{2})$ corresponds to the complex gain of the $l$\textsuperscript{th} multipath ray in the $i$\textsuperscript{th} cluster and $\sigma_{\alpha,i}^{2}$ is the average power of the $i$\textsuperscript{th} cluster, such that $\sum_{i=1}^{N_{\mathrm{cl}}} \sigma_{\alpha,i}^{2} = \sqrt{N_\mathrm{r} N_\mathrm{t} / N_{\mathrm{cl}} N_{\mathrm{ray}}}$ to ensures $\mathbb{E} [\| \mathbf{H} \|^2_\mathrm{F}] = N_\mathrm{r} N_\mathrm{t}$; 
		$\mathbf{a}_\mathrm{r}(\phi_{il}^{\mathrm{r}},\theta_{il}^{\mathrm{r}})$ and $\mathbf{a}_\mathrm{t} (\phi_{il}^{\mathrm{t}},\theta_{il}^{\mathrm{t}})$ are the array response vectors of the receiver and transmitter, respectively; 
		$\phi_{il}^{\mathrm{t}}$ and $\theta_{il}^{\mathrm{t}}$ are the azimuth and elevation angles of departure (AoD), and $\phi_{il}^{\mathrm{r}}$ and $\theta_{il}^{\mathrm{r}}$ are the azimuth and elevation angles of arrival (AoA), which are modeled as Laplacian distributed random variable, with mean $\phi_{i}^{\mathrm{t}},\theta_{i}^{\mathrm{t}},\phi_{i}^{\mathrm{r}},\theta_{i}^{\mathrm{r}}$ uniformly-distributed over $[-\pi,\pi)$, and angular spread of $\sigma_{\phi}^{\mathrm{t}},\sigma_{\theta}^{\mathrm{t}},\sigma_{\phi}^{\mathrm{r}}, \sigma_{\theta}^{\mathrm{r}}$~\cite{Ayach2014,Sohrabi2016}.	
		We further assume a $\sqrt{N} \times \sqrt{N}$ uniform square planar array~(USPA), such that the array response vector is
		\vspace{-0.1cm}
		\begin{equation}
		\vspace{-0.1cm}
		\begin{split}
		\mathbf{a} (\phi,\theta) = &  \frac{1}{\sqrt{N}} \left[ 1, \cdots,  e^{j\frac{2\pi d}{\lambda} \left[ h \sin(\phi) \sin(\theta) + v \cos(\theta) \right]}, \right.\\
		& \left. \cdots,  e^{j\frac{2\pi d}{\lambda} \left[ (\sqrt{N}-1) \sin(\phi) \sin(\theta) + (\sqrt{N}-1) \cos(\theta) \right] }  \right]^\mathrm{T}
		\end{split}		
		\end{equation}
		where \mbox{$0 \leq h < \sqrt{N}-1$} and \mbox{$0 \leq v < \sqrt{N}-1$} are the indexes of the antenna element in the 2D plane, $d$ is the spacing between elements, and $\lambda$ is the signal wavelength.
		
		Here, we seek to design the hybrid precoder and combiner to maximize the average achievable sum-rate over all subcarriers. Such a problem is formulated as
		\vspace{-0.05cm}			
		\begin{equation}			
			\label{eq:prob_statement}
			\begin{aligned}
			& \underset{\mathbf{F}_{\mathrm{RF}}, \mathbf{F}_{\mathrm{BB},m},  \mathbf{W}_{\mathrm{RF}}, \mathbf{W}_{\mathrm{BB},m}}{\text{ max}} 
			& & \hspace{-0.2cm} \frac{1}{M} \sum_{m=1}^{M} R_\mathrm{sum,m} \\
			& \text{~~~~~~~~~~~s.t.}
			& & \hspace{-0.2cm} \mathbf{F}_{\mathrm{RF}} \in \mathcal{F}_{\mathrm{RF}} \\
			& 
			& & \hspace{-0.2cm} \mathbf{W}_{\mathrm{RF}} \in \mathcal{W}_{\mathrm{RF}} \\
			&&& \hspace{-0.2cm} \left\| \mathbf{F}_{\mathrm{RF}}  \mathbf{F}_{\mathrm{BB},m} \right\|_{\mathrm{F}}^2 = N_\mathrm{s}, ~~ 1 \leq m \leq M
			\end{aligned}
			\vspace{-0.05cm}
		\end{equation}
		where $\mathcal{F}_{\mathrm{RF}}$ and $\mathcal{W}_{\mathrm{RF}}$ are, respectively, the set of all feasible analog precoders and combiners (i.e., all $N_\mathrm{t} \times N_\mathrm{t}^{\mathrm{RF}}$ and $N_\mathrm{r} \times N_\mathrm{r}^{\mathrm{RF}}$ matrices with constant modulus entries) and $\| \mathbf{F}_{\mathrm{RF}} \mathbf{F}_{\mathrm{BB},m} \|_{\mathrm{F}}^2~=~N_\mathrm{s}$ is the per-subcarrier total power constraint~\cite{Sohrabi2017}. 				
		For simplicity, we assume $N_\mathrm{t}^{\mathrm{RF}} = N_\mathrm{r}^{\mathrm{RF}} = N_{\mathrm{s}}$. 
		Note that solving this problem is very challenging as it requires a joint optimization over multiple matrix variables subjected to the non-convex constant modulus constraint imposed by the phase-shifters on analog precoders and combiners.
		The optimal solution to \eqref{eq:prob_statement} is obtained by finding the hybrid precoders and combiners, $\mathbf{F}_{m}$ and $\mathbf{W}_{m}$, that diagonalize the effective channel $\mathbf{W}_{m}^{\mathrm{H}} \mathbf{H}_{m} \mathbf{F}_{m}$. 
		For fully-digital beamforming, this is achieved by setting the precoders and combiners as the singular vectors associated with the $N_\mathrm{s}$ largest singular values of the channel matrices $\mathbf{H}_{m}$. 
		While narrowband HBF designs usually seek to approximate the optimal fully-digital solution, this is very challenging for OFDM HBF design since all subcarriers share the analog precoders and combiners.
		
\section{Proposed Hybrid Beamforming Design}
\label{sec:OFDM_hbf_design}

	Here, we propose a novel joint hybrid precoder and combiner design for OFDM systems. The analog precoder and combiner are designed by the proposed low-rank constrained Tucker2 decomposition, which aims to  maximize the sum of the baseband effective channel gains over all subcarriers, while suppressing interference among data streams within the same subcarrier by seeking orthogonality among the analog precoders/combiners of different data streams. Such orthogonality, which ensures the diagonalization of the effective channel, cannot be completely attained by the analog beamforming alone. Therefore, we design the digital precoder and combiner to further attain the complete diagonalization.
	In this section, we briefly describe the Tucker decomposition, formulate the OFDM analog beamforming design problem as a low-rank constrained Tucker2 decomposition and propose an algorithm for solving it, present the digital beamforming design, and finally, analyze the computational complexity of the proposed OFDM HBF design.
	
	\vspace{-0.2cm}
	\subsection{Tucker Decomposition}
	
	The Tucker decomposition decomposes a tensor into a core tensor multiplied by a factor matrix along each mode.	
	The truncated rank-$(R_1,R_2,R_3)$ Tucker decomposition of a 3-way tensor $\mathcal{X} \in \mathbb{C}^{I_1 \times I_2 \times I_3}$ is defined as~\cite{Kolda2009}
	\vspace{-0.1cm}
	\begin{equation}			
		\label{eq:tucker_decomposition}
		\hspace{-0.01cm} \mathcal{X} \hspace{-0.06cm} \approx  \hspace{-0.05cm} \mathcal{G} \hspace{-0.05cm} \times_1 \hspace{-0.05cm} \mathbf{A}^{(1)} \hspace{-0.05cm} \times_2 \hspace{-0.05cm} \mathbf{A}^{(2)} \hspace{-0.05cm} \times_3 \hspace{-0.05cm} \mathbf{A}^{(3)} \hspace{-0.06cm} = \hspace{-0.12cm} \sum_{r_1=1}^{R_1} \sum_{r_2=1}^{R_2} \sum_{r_3=1}^{R_3} \hspace{-0.06cm} g_{r_1 r_2 r_3} \mathbf{a}_{r_1}^{(1)} \hspace{-0.02cm} \circ \hspace{-0.00cm} \mathbf{a}_{r_2}^{(2)} \hspace{-0.02cm} \circ \hspace{-0.00cm} \mathbf{a}_{r_3}^{(3)} \hspace{-0.2cm}
		\vspace{-0.1cm}
	\end{equation}
	where $\mathbf{A}^{(1)} \in \mathbb{C}^{I_1 \times R_1}$, $\mathbf{A}^{(2)} \in \mathbb{C}^{I_2 \times R_2}$, and $\mathbf{A}^{(3)} \in \mathbb{C}^{I_3 \times R_3}$ are the factor matrices (which are usually orthogonal), $\mathcal{G} \in \mathbb{C}^{R_1 \times R_2 \times R_3}$ is the core tensor, $\circ$ is the outer product, and $\times_n$ is the $n$-mode product\footnote{For details on tensor notations and operations, please refer to~\cite{Kolda2009}.}.
	The  decomposition in \eqref{eq:tucker_decomposition} is obtained by solving the following problem:
	\vspace{-0.1cm}
	\begin{equation}			
		\label{eq:tucker_decomposition_problem_min}
		\begin{aligned}
		& \underset{\mathcal{G}, \mathbf{A}^{(1)}, \mathbf{A}^{(2)}, \mathbf{A}^{(3)}}{\text{min}} 
		& & \hspace{-0.2cm} \left\| \mathcal{X} - \mathcal{G} \times_1 \mathbf{A}^{(1)} \times_2 \mathbf{A}^{(2)} \times_3 \mathbf{A}^{(3)} \right\|^{2}_{\mathrm{F}}  \\
		& \text{~~~~~~s.t.}
		& & \hspace{-0.2cm} \mathcal{G} \in \mathbb{C}^{R_1 \times R_2 \times R_3}\\
		&
		& & \hspace{-0.2cm} {\mathbf{A}^{(n)}}^\mathrm{H} \mathbf{A}^{(n)} = \mathbf{I}_{R_n},~\forall n\in {1,2,3}.
		\end{aligned}
		\vspace{-0.1cm}
	\end{equation}
	
	The solution of \eqref{eq:tucker_decomposition_problem_min} can be obtained by the alternating least squares (ALS) method, in which we alternately solve the problem for one factor matrix, while fixing the others. Thereby, the solution for the factor matrix $\mathbf{A}^{(n)}$ for $n \in \{1,2,3\}$ is obtained by solving~\cite{Kolda2009,Lathauwer2000,Lathauwer2000b}
	\vspace{-0.1cm}
	\begin{equation}			
		\label{eq:tucker_decomposition_ALS}
		\begin{aligned}
		& \underset{\mathbf{A}^{(n)}}{\text{max}} 
		& & \hspace{-0.2cm} \left\| \mathcal{X} \times_1 \mathbf{A}^{(1)} \times_2 \mathbf{A}^{(2)} \times_3 \mathbf{A}^{(3)} \right\|^{2}_{\mathrm{F}}  \\
		& \text{~~~s.t.}
		& & \hspace{-0.2cm} {\mathbf{A}^{(n)}}^\mathrm{H} \mathbf{A}^{(n)} = \mathbf{I}_{R_n}\\
		\end{aligned}
		\vspace{-0.1cm}
	\end{equation}
	where the cost function can be written as $ \| {\mathbf{A}^{(n)}}^\mathrm{H} \mathbf{Z} \|^{2}_{\mathrm{F}}$,
	with $\mathbf{Z} = \mathbf{X}_{(n)} ( \mathbf{A}^{(N)} \otimes \cdots \otimes \mathbf{A}^{(n+1)} \otimes \mathbf{A}^{(n-1)} \otimes \cdots \mathbf{A}^{(1)} )$, and $\mathbf{X}_{(n)}$ denoting the mode-n matricization of tensor $\mathcal{X}$~\cite{Kolda2009,Lathauwer2000,Lathauwer2000b}. It is worth noticing that  Tucker decompositions are not unique and that the ALS solution is not guaranteed to converge to the global optimum of \eqref{eq:tucker_decomposition_problem_min}, but only to points where the objective function cease to decrease~\cite{Kolda2009}.
	For the HBF problem, we are interested in a particular case of the Tucker decomposition, termed the Tucker2 decomposition, for which one of the factor matrix is set to the identity matrix.

	\vspace{-0.2cm}	
	\subsection{Analog Precoder and Combiner Design}
	
		In order to design the analog precoder and combiner, we propose a constrained Tucker2 decomposition. We arrange the channel tensor $\mathcal{H} \in \mathbb{C}^{N_r \times N_t \times M}$, such that each subcarrier's channel matrix corresponds to a frontal slice of $\mathcal{H}$, and seek to find the constant modulus constrained orthogonal analog precoder and combiner matrices that best approximate the channel tensor.
		This problem is written as	
		\vspace{-0.1cm}			
		\begin{equation}			
		\label{eq:Channel_tucker_decomposition}
		\begin{aligned}
		& \underset{\mathbf{A}^{(n)}}{\text{max}}  
		& & \hspace{-0.2cm} \left\| \mathcal{H} \times_1 \mathbf{W}_{\mathrm{RF}} \times_2 \mathbf{F}_{\mathrm{RF}} \times_3 \mathbf{I}_M \right\|^{2}_{\mathrm{F}}  \\
		& \text{~~~s.t.}
		& & \hspace{-0.2cm} \mathbf{F}_{\mathrm{RF}} \in \mathcal{F}_{\mathrm{RF}},~~ \mathbf{F}_{\mathrm{RF}}^\mathrm{H} \mathbf{F}_{\mathrm{RF}} = \mathbf{I}_{N_s}\\
		& 
		& & \hspace{-0.2cm} \mathbf{W}_{\mathrm{RF}} \in \mathcal{W}_{\mathrm{RF}},~~ \mathbf{W}_{\mathrm{RF}}^\mathrm{H} \mathbf{W}_{\mathrm{RF}} = \mathbf{I}_{N_s}. \\
		\end{aligned}
		\vspace{-0.15cm}
		\end{equation}
		However, due to the constant modulus constraint, finding orthogonal analog precoder and combiner matrices can lead to an intractable problem. 
		
		To handle this issue, we break down problem~\eqref{eq:Channel_tucker_decomposition} into $N_\mathrm{s}$ successive rank-$(1,1)$ approximation sub-problems.
		Each sub-problem corresponds to designing the analog precoder and combiner vector pair for one data stream in all subcarriers. These sub-problems are formulated as
		\vspace{-0.2cm}	
		\begin{equation}	
			\label{eq:OFDM_analogBF}
			\begin{aligned}
				& \underset{\mathbf{w}_{i}, \mathbf{f}_{i}}{\text{max}} 
				& & \hspace{-0.2cm} \left\| \mathcal{H}_\mathrm{res}^{(i)} \times_1 \mathbf{w}_{i} \times_2 \mathbf{f}_{i} \times_3 \mathbf{I}_M \right\|^{2}_{\mathrm{F}} \\
				& \text{~~s.t.}
				& & \hspace{-0.2cm} \left|  [\mathbf{f}_{i}]_{m} \right| =  1/\sqrt{N_\mathrm{t}},~~ \mathbf{f}_{i}^\mathrm{H} \mathbf{f}_{i} = 1\\
				& 
				& & \hspace{-0.2cm} \left|  [\mathbf{w}_{i}]_{m} \right| =  1/\sqrt{N_\mathrm{r}},~~ \mathbf{w}_{i}^\mathrm{H} \mathbf{w}_{i} = 1\\
			\end{aligned}
			\vspace{-0.15cm}	
		\end{equation}
		where $\mathbf{F}_{\mathrm{RF}} = [\mathbf{f}_{1} \cdots \mathbf{f}_{N_\mathbf{s}}]$ and $\mathbf{W}_{\mathrm{RF}} = [\mathbf{w}_{1} \cdots \mathbf{w}_{N_\mathbf{s}}]$.
		Note that, although we have relaxed the orthogonality constraints to ensure the tractability of the problem, these constraints are reinforced by using the residual channel tensor $\mathcal{H}_{\mathrm{res}}^{(i)}$, instead of the original channel tensor $\mathcal{H}$.
		To solve problem \eqref{eq:OFDM_analogBF}, we recall to the ALS solution by alternately and iteratively solving 
		\vspace{-0.25cm}	
		\begin{equation}	
			\label{eq:OFDM_analogBF_w}
			\begin{aligned}
				& \underset{\mathbf{w}_{i}}{\text{max}} 
				& & \hspace{-0.2cm} \left\| \mathbf{w}_{i}^{\mathrm{H}} {\mathbf{H}^{(i)}_\mathrm{res}}_{(1)} \left( \mathbf{I}_M \otimes \mathbf{f}_{i}\right) \right\|^{2}_{\mathrm{F}} \\
				& \text{~~s.t.}
				& & \hspace{-0.2cm} \left|  [\mathbf{w}_{i}]_{m} \right| =  1/\sqrt{N_\mathrm{r}}, ~~\mathbf{w}_{i}^\mathrm{H} \mathbf{w}_{i} = 1
			\end{aligned}
			\vspace{-0.1cm}	
		\end{equation}
		and
		\vspace{-0.1cm}	
		\begin{equation}	
			\label{eq:OFDM_analogBF_f}
			\begin{aligned}
				& \underset{\mathbf{f}_{i}}{\text{max}} 
				& & \hspace{-0.2cm} \left\| \mathbf{f}_{i}^{\mathrm{H}} {\mathbf{H}^{(i)}_\mathrm{res}}_{(2)} \left( \mathbf{I}_M \otimes \mathbf{w}_{i} \right) \right\|^{2}_{\mathrm{F}} \\
				& \text{~~s.t.}
				& & \hspace{-0.2cm} \left|  [\mathbf{f}_{i}]_{m} \right| =  1/\sqrt{N_\mathrm{t}}, ~~ \mathbf{f}_{i}^\mathrm{H} \mathbf{f}_{i} = 1.
			\end{aligned}
			\vspace{-0.15cm}	
		\end{equation}
	
		Problems \eqref{eq:OFDM_analogBF_w} and \eqref{eq:OFDM_analogBF_f} are still non-convex problems due to the constant modulus constraint. 
		To obtain feasible solutions, we solve their convex relaxations, by dropping the constant modulus constraint. The relaxed solutions to problems \eqref{eq:OFDM_analogBF_w} and \eqref{eq:OFDM_analogBF_f} are, respectively, given by the principal eigenvector of ${\mathbf{H}^{(i)}_\mathrm{res}}_{(1)} ( \mathbf{I}_M \otimes \mathbf{f}_{i} {\mathbf{f}_{i}}^{\mathrm{H}})  {\mathbf{H}^{(i)}_\mathrm{res}}_{(1)}^{\mathrm{H}}$ and ${\mathbf{H}^{(i)}_\mathrm{res}}_{(2)} ( \mathbf{I}_M \otimes \mathbf{w}_{i} {\mathbf{w}_{i}}^{\mathrm{H}} ) {\mathbf{H}^{(i)}_\mathrm{res}}_{(2)}^{\mathrm{H}}$. Solving these eigenproblems for each ALS iteration is computationally expensive, therefore, we approximate the solution by a single iteration of the power-iteration method, which yields
		\vspace{-0.2cm}	
		\begin{equation}	
		\label{eq:OFDM_analogBF_solution_w}
		\hat{\mathbf{w}}_{i} =  {\mathbf{H}^{(i)}_\mathrm{res}}_{(1)} \left( \mathbf{I}_M \otimes \mathbf{f}_{i} {\mathbf{f}_{i}}^{\mathrm{H}}\right)  {\mathbf{H}^{(i)}_\mathrm{res}}_{(1)}^{\mathrm{H}}  \mathbf{w}_{i}
		\vspace{-0.2cm}	
		\end{equation}
		and
		\begin{equation}	
		\label{eq:OFDM_analogBF_solution_f}
		\hat{\mathbf{f}_{i}} = {\mathbf{H}^{(i)}_\mathrm{res}}_{(2)} \left( \mathbf{I}_M \otimes \mathbf{w}_{i} {\mathbf{w}_{i}}^{\mathrm{H}} \right) {\mathbf{H}^{(i)}_\mathrm{res}}_{(2)}^{\mathrm{H}}  \mathbf{f}_{i}.
		\vspace{-0.12cm}	
		\end{equation}
		%
		The relaxed solutions in \eqref{eq:OFDM_analogBF_solution_w} and \eqref{eq:OFDM_analogBF_solution_f} are then projected onto the set of constant modulus vectors, by extracting their phases through $\mathbf{w}_{i} \hspace{-0.1cm} = \hspace{-0.1cm}  (1/\sqrt{N_\mathrm{r}}) \psi \lbrace \mathbf{\mathbf{\hat{w}}_{i}} \rbrace$ and $\mathbf{f}_{i} \hspace{-0.1cm}  =  \hspace{-0.1cm}  (1/\sqrt{N_\mathrm{t}}) \psi \lbrace \mathbf{\mathbf{\hat{f}}_{i}} \rbrace$, respectively, where $ \psi \lbrace \mathbf{v} \rbrace = e^{j\angle{\mathbf{\mathbf{v}}}}$~\cite{Zilli2020}.
		After computing the analog precoder and combiner vector pair, we update the residual channel matrix  to remove the contribution of such a pair by setting
		\vspace{-0.2cm}
		\begin{equation}
		\label{eq:residualTensorUpdate}
		\mathbf{H}^{(i+1)}_{\mathrm{res}~(1)} = \mathbf{P}_\mathrm{w_i} {\mathbf{H}_{\mathrm{res}}^{(i)}}_{(1)} (\mathbf{I}_M \otimes \mathbf{P}_\mathrm{f_i})
		\vspace{-0.15cm}
		\end{equation}
		where $\mathbf{P}_\mathrm{w_i} = [ \mathbf{I}_{N_\mathrm{r}} - \mathbf{w}_{i} \mathbf{w}_{i}^{\mathrm{H}} ]$ and $\mathbf{P}_\mathrm{f_i} = [ \mathbf{I}_{N_\mathrm{t}} - \mathbf{f}_{i} \mathbf{f}_{i}^{\mathrm{H}}]$. This procedure is repeated until all the $N_\mathrm{s}$ analog precoder and combiner vector pairs have been designed.
		Note that using the single-power-iteration approximations in~\eqref{eq:OFDM_analogBF_solution_w}~and~\eqref{eq:OFDM_analogBF_solution_f} instead of the eigenproblem solutions has almost no impact on the system performance; although, the proposed algorithm may require few more ALS iterations.
		The proposed analog beamforming design is summarized in Algorithm~\ref{alg:ODFM-CSVD-HFB}. 
	
		\begin{center}
			\begin{algorithm}[t]
				\renewcommand{\baselinestretch}{1}\normalsize
				\SetAlgoLined
				\SetArgSty{textrm} 
				\DontPrintSemicolon		
				$\text{Initialize } \mathbf{F}_{\mathrm{RF}} = [\;\;] \text{; } \mathbf{W}_{\mathrm{RF}} = [\;\;] \text{; } \mathcal{H}_{\mathrm{res}}^{(1)} = \mathcal{H}$ \;
				\For{$i = 1:N^{\mathrm{s}}$}{
					$ \text{Initialize } \mathbf{w}_{i} \text{ and } \mathbf{f}_{i} \text{ randomly and } \eta=0$ \;
					$ \text{Initialize }  \delta^{(0)}  = 0 \text{ and } \delta^{(1)} = \frac{ \| \mathcal{H}_\mathrm{res}^{(i)} \times_1 \mathbf{w}_{i} \times_2 \mathbf{f}_{i}  \times_3 \mathbf{I}_M \|^{2}_{\mathrm{F}} }{M}$\;
					\While{$\| \delta^{(\eta + 1)} - \delta^{(\eta)} \|^2  \geq \varepsilon \text{ and } \eta \leq N_{\mathrm{ite}}$}{
						$\eta = \eta + 1 $ \;
						$\mathbf{w}_{i} = \frac{1}{\sqrt{N_\mathrm{t}}} \psi \left\lbrace {\mathbf{H}^{(i)}_\mathrm{res}}_{(1)} \left( \mathbf{I}_M \otimes \mathbf{f}_{i}  {\mathbf{f}_{i}}^{\mathrm{H}}\right)  {\mathbf{H}^{(i)}_\mathrm{res}}_{(1)}^{\mathrm{H}}  \mathbf{w}_{i} \right\rbrace$ \;
						$\mathbf{f}_{i} =  \frac{1}{\sqrt{N_\mathrm{t}}} \psi \left\lbrace  {\mathbf{H}^{(i)}_\mathrm{res}}_{(2)} \left( \mathbf{I}_M \otimes \mathbf{w}_{i}  {\mathbf{w}_{i}}^{\mathrm{H}} \right) {\mathbf{H}^{(i)}_\mathrm{res}}_{(2)}^{\mathrm{H}}  \mathbf{f}_{i} \right\rbrace $ \;						
						$ \delta^{(\eta+1)} = \frac{ \| \mathcal{H}_\mathrm{res}^{(i)} \times_1 \mathbf{w}_{i} \times_2 \mathbf{f}_{i} \times_3 \mathbf{I}_M \|^{2}_{\mathrm{F}}  }{M}$ \;
					}
					$\text{Update } \mathcal{H}_{\mathrm{res}}^{(i+1)} \text{ according to~\eqref{eq:residualTensorUpdate} }$ \;
					$\mathbf{F}_{\mathrm{RF}} = [\; \mathbf{F}_{\mathrm{RF}} \quad \mathbf{f}_{i} \;]  \text{ and } \mathbf{W}_{\mathrm{RF}} = [\; \mathbf{W}_{\mathrm{RF}} \quad \mathbf{w}_{i} \;]  $ \;
				}		
				\Return{$\mathbf{F}_{\mathrm{RF}}, \mathbf{W}_{\mathrm{RF}}$ }
				\caption{OFDM Analog Beamforming Design}
				\label{alg:ODFM-CSVD-HFB}
			\end{algorithm}	
		\end{center}
		\setlength{\textfloatsep}{2pt}
		\vspace{-1.0cm}	
		
	\vspace{-0.2cm}
	\subsection{Digital Precoder and Combiner Design}
		\vspace{-0.05cm}
	
		The optimal digital precoder and combiner is obtained on a per-subcarrier basis through the effective channel's SVD. 
		The effective channel of the $m^{\mathrm{th}}$ subcarrier, $\mathbf{H}_{m}^{\mathrm{eff}} = \mathbf{W}_{\mathrm{RF}}^{\mathrm{H}} \mathbf{H}_{m} \mathbf{F}_{\mathrm{RF}}$, takes into account the effect the analog precoder and combiner and has SVD as $\mathbf{H}^{\mathrm{eff}}_{m} = \mathbf{U}_{\mathrm{e}} \mathbf{\Sigma}_{\mathrm{e}} \mathbf{V}_{\mathrm{e}}^{\mathrm{H}}$.
		The digital precoder and combiner are designed by taking the $N_\mathrm{s}$ column vectors of $\mathbf{V}_{\mathrm{e}}$ and $\mathbf{U}_{\mathrm{e}}$ associated with the $N_\mathrm{s}$ largest singular values, i.e., $\mathbf{F}_{\mathrm{BB},m} = \mathbf{V}_{\mathrm{e}} [~:~,1:N_\mathrm{s}]$ and $\mathbf{W}_{\mathrm{BB},m} = \mathbf{U}_{\mathrm{e}} [~:~,1:N_\mathrm{s}]$.
		Finally, we normalize the digital precoder matrix to ensure the per-subcarrier total transmitted power constraint by making 
		\vspace{-0.1cm}
		\begin{equation}
			\mathbf{F}_{\mathrm{BB},m} = \sqrt{N_\mathrm{s}} \frac{\mathbf{F}_{\mathrm{BB},m}}{\| \mathbf{F}_{\mathrm{RF}} \mathbf{F}_{\mathrm{BB},m} \|_{\mathrm{F}}}.
			\vspace{-0.25cm}
		\end{equation}

	\vspace{-0.2cm}	
	\subsection{Computational complexity}
		\vspace{-0.05cm}
		\label{sec:sub:OFDM_complexity}
		
		\begin{table}[t]
			\vspace{-0.2cm}	
			\caption{Computational Complexity of OFDM HBF Design Methods}		
			\renewcommand{\arraystretch}{1.3}
			\centering			
			\label{tab:OFDM_complexity}
			\vspace{-0.15cm}			
			\begin{tabular}{c|c}	
				\hline		
				\textbf{Method} & \textbf{Computational complexity order} \\ \hline
				Proposed Method & \hspace{-0.1cm} $M [N_\mathrm{s}^3 + N_\mathrm{s}^2 N_\mathrm{t} + N_\mathrm{s}  N_\mathrm{ite} ( N_\mathrm{t} N_\mathrm{r} +  N_{\mathrm{t}}^2)]$ \\ \hline
				PE-AltMin~\cite{Yu2016} & \hspace{-0.1cm} $ M  N_\mathrm{ite} [ N_\mathrm{s}^3 + N_\mathrm{s}^2(N_\mathrm{t} + N_\mathrm{r}) ] + M N_\mathrm{s} N_\mathrm{t} N_\mathrm{r} $ \\ \hline
				\mbox{HBF-LSAA}~\cite{Sohrabi2017} &  \hspace{-0.1cm} $\left.  N_\mathrm{ite} [ N_\mathrm{s}^4 + N_\mathrm{s}^3(N_\mathrm{t} + N_\mathrm{r})  + N_\mathrm{s}^2 (N_\mathrm{t}^2  N_\mathrm{r}^2)  ] \right. $ \\ 
				& $\left. + M [N_\mathrm{s}^3 + N_\mathrm{t}^2 N_\mathrm{r} + N_\mathrm{r}^2  N_\mathrm{s}] \right. $  \\ \hline
				SS-SVD~\cite{Tsai2019} & \hspace{-0.1cm} $\left.  M[N_\mathrm{s}^3 + N_\mathrm{t}^2 N_\mathrm{r} + (N_\mathrm{r}^2 + N_\mathrm{s}^2) N_\mathrm{t}]  + N_\mathrm{s} (N_\mathrm{t}^2 +  N_\mathrm{r}^2) \right. $ \\ \hline
				ICSI-HBF~\cite{Chiang2018} & \hspace{-0.1cm} $M [N_\mathrm{s}^3 + N_\mathrm{s}^2 (N_\mathrm{t} + N_\mathrm{r})] + C^2 N_\mathrm{s}$  \\ 
				\hline
			\end{tabular}
		\vspace{-0.3cm}
		\end{table}
		\setlength{\textfloatsep}{12pt}
		
		The complexity order of the proposed method and that in~\cite{Yu2016,Sohrabi2017},~\cite{Tsai2019}, and~\cite{Chiang2018} are shown in Table~I, where $C$ represents the cardinality of the codebook in~\cite{Chiang2018}.
		Specifically, the computational complexity of Alg.~\ref{alg:ODFM-CSVD-HFB} is approximately $\mathcal{O}\lbrace  M N_\mathrm{s} N_\mathrm{ite} (N_\mathrm{t} N_\mathrm{r} + N_{\mathrm{t}}^2)\rbrace$, where $N_\mathrm{ite}$ is the maximum number of iterations to compute an analog precoder and combiner vector pair.
		The digital beamforming design requires computing the baseband effective channel, its SVD, and the digital precoder normalization for every subcarrier. These operations require $\mathcal{O} \left\lbrace M (N_\mathrm{s}^3 +  N_\mathrm{s}^2 N_\mathrm{t} + N_\mathrm{s} N_\mathrm{t} N_\mathrm{r}) \right\rbrace$.
		Therefore, the overall complexity of the proposed design is $\mathcal{O}\left\lbrace M [N_\mathrm{s}^3 + N_\mathrm{s}^2 N_\mathrm{t} + N_\mathrm{s}  N_\mathrm{ite} ( N_\mathrm{t} N_\mathrm{r} + N_{\mathrm{t}}^2)]\right\rbrace$.
		Note that the complexity is mostly associated to the computations in line 7-9 of Alg.~\ref{alg:ODFM-CSVD-HFB}. Nevertheless, we can exploit the structure of those operations to implement them in a parallel fashion, thereby reducing significantly the computation time.

\vspace{-0.2cm}		
\section{Simulation Results}
	\label{sec:simulations}

	In this section, we present simulation results to evaluate the performance of the proposed algorithms and compare with alternative designs.
	We assume a mmWave channel with $N_\mathrm{cl} = 5$ and $N_\mathrm{ray} = 10$, angular spread $\sigma_{\phi}^{\mathrm{t}}=\sigma_{\theta}^{\mathrm{t}}=\sigma_{\phi}^{\mathrm{r}}= \sigma_{\theta}^{\mathrm{r}} = 10^{\circ}$, and a USPA with antenna spacing $d = \lambda / 2$~\cite{Ayach2014,Yu2016}. The signal-to-noise ratio is defined as $\text{SNR} = \frac{\rho}{\sigma_\mathrm{n}^2}$~\cite{Ayach2014}.
	The proposed design is compared with the optimal unconstrained HBF, the PE-AltMin~\cite{Yu2016}, \mbox{HBF-LSAA}~\cite{Sohrabi2017}, SS-SVD~\cite{Tsai2019}, and ICSI-HBF~\cite{Chiang2018} (for which we used a $C=64$ orthogonal beam codebook). In all experiments, we consider an OFDM system with 1024 subcarriers and adopt $\varepsilon = 1$ and $N_{\mathrm{ite}}=10$ to control the convergence of Alg.~1. The results are obtained by averaging the achievable sum-rate over $1000$ channel realizations. 
		
\begin{figure}[!t]
		\centering
		\includegraphics[width=.75\columnwidth]{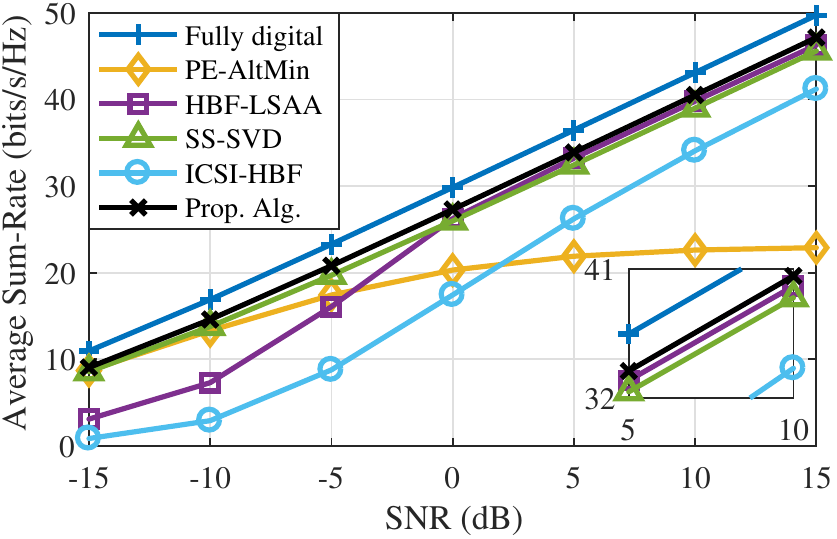}
		\vspace{-0.1cm}
		\caption{\small{Average sum-rate vs. SNR for $N_\mathrm{t} = N_\mathrm{r} = 64$, $M=1024$, and $N_\mathrm{t}^{\mathrm{RF}} = N_\mathrm{r}^{\mathrm{RF}} = N_\mathrm{s} = 4$.}}
		\vspace{-0.3cm}
		\label{fig:OFDM_example1_sr}
	\end{figure}

	First, we evaluate the average sum-rate versus the SNR of a system with $N_\mathrm{t} = N_\mathrm{r} = 64$ antennas, arranged in a $8 \times 8$ array configuration,	and $N_\mathrm{t}^{\mathrm{RF}} = N_\mathrm{r}^{\mathrm{RF}} = 4$ RF chains communicating through $N_\mathrm{s} = 4$ data streams per subcarrier. The results, depicted in Fig.~\ref{fig:OFDM_example1_sr}, show that the proposed design outperforms other designs over all SNR range. Notice  that the PE-AltMin and \mbox{HBF-LSAA} have their average sum-rate deteriorating as the SNR increases and decreases, respectively.
			
\begin{figure}[!t]
		\centering
		\includegraphics[width=.75\columnwidth]{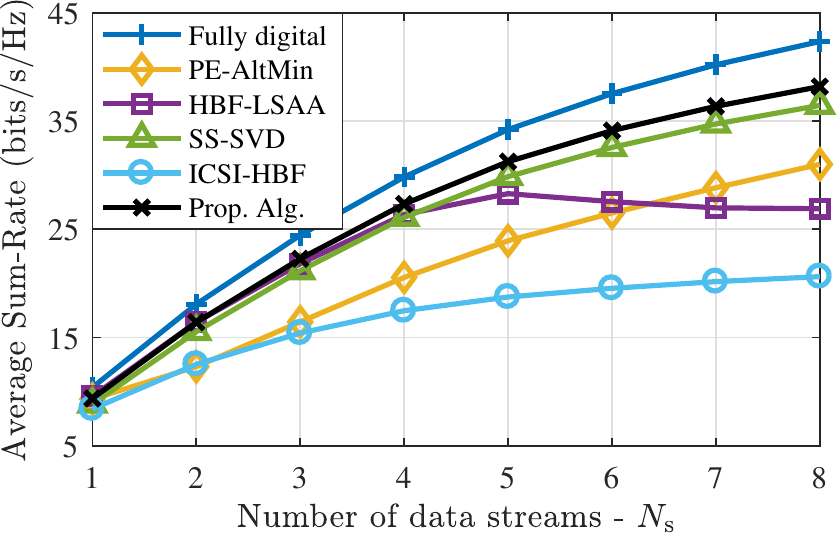}
		\vspace{-0.1cm}
		\caption{\small{Average sum-rate vs. the number of data streams for $N_\mathrm{t} = N_\mathrm{r} = 64$, $M=1024$, and SNR~=~0~dB.}}
		\vspace{-0.3cm}
		\label{fig:OFDM_example2}
	\end{figure}

	Second, we evaluate the average sum-rate versus the number of data streams per subcarrier. We assume a system with $N_\mathrm{t} = N_\mathrm{r} = 64$ antennas ($8 \times 8$ array) and $N_\mathrm{t}^{\mathrm{RF}} = N_\mathrm{r}^{\mathrm{RF}} = N_\mathrm{s}$ RF chains operating at SNR~=~0~dB. The results, presented in Fig.~\ref{fig:OFDM_example2}, show that the proposed method outperforms the other methods over all range of $N_\mathrm{s}$. Also, note that the average sum-rate of the \mbox{HBF-LSAA} method starts to decrease once $N_\mathrm{s}\geq4$.
			
\begin{figure}[!t]
		\centering
		\includegraphics[width=.75\columnwidth]{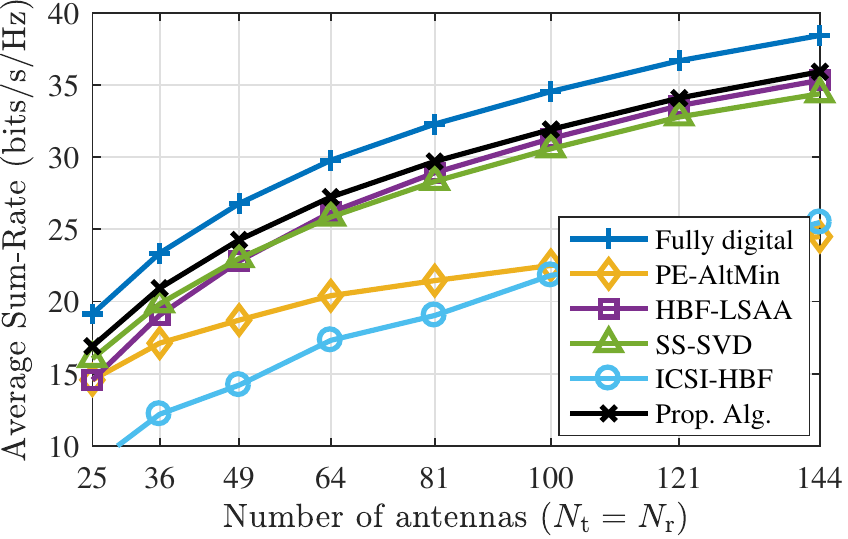}
		\vspace{-0.1cm}
		\caption{\small{Average sum-rate vs. the number of antennas $N_\mathrm{t} = N_\mathrm{r}$ for SNR~=~0~dB, $M=1024$ and $N_\mathrm{s} = 4$.}}
		\vspace{-0.2cm}
		\label{fig:OFDM_example3}
	\end{figure}

	Finally, we evaluate the average sum-rate versus the number of antennas $N_\mathrm{t} = N_\mathrm{r}$ for a system with $N_\mathrm{t}^{\mathrm{RF}} = N_\mathrm{r}^{\mathrm{RF}} = 4$ RF chains, communicating through $N_\mathrm{s} = 4$ data streams in each subcarrier at a SNR~=~0~dB.
	Results are depicted in Fig.~\ref{fig:OFDM_example3}, which show that the proposed design is also able to outperform the other designs over a wide range of array sizes.
	Here, notice that the average sum-rate attained by the \mbox{HBF-LSAA} method tends to outperforms the proposed design when the number of antennas grows. This is reasonable, since that method was originally conceived for asymptotically large arrays.
	For practical array sizes, however, the proposed design attains average sum-rates superior to the \mbox{HBF-LSAA} method.

	The proposed HBF design attains near-optimal performance over a wide range SNR, number of data streams, and number of antennas, outperforming other methods in all simulation scenarios. 
	The higher performance of the proposed design may be attributed to the joint computation of the analog precoder and combiner vector pairs. While other methods compute the analog precoders and combiners separately, such as those in~\cite{Yu2016}~and~\cite{Tsai2019}, or may use the information of the precoder to compute the combiner, as in~\cite{Sohrabi2017}, the proposed design uses an alternate optimization approach that ensures the analog precoder and combiner are designed together and optimized to each other.
	The results have also revealed limitations of the PE-AltMin and HBF-LSAA designs in terms of their achieved sum-rate, respectively, in high and low SNR scenarios. The HBF-LSAA design has also shown a performance degradation when increasing the number of data streams.

	It should be mentioned that as for the unconstrained tensor decompositions in~\cite{Kolda2009} and \cite{Lathauwer2000}, there is no guarantee that the proposed constrained Tucker2 decomposition is unique or that it will converge to a global minimum. In fact, through extensive simulations, we have been able to verify that the non-uniqueness is, in general, up to a permutation and/or a phase rotation in the columns of $\mathbf{F}_{\mathrm{RF}}$ and $\mathbf{W}_{\mathrm{RF}}$.
	Simulations under different scenarios have also shown that each inner iteration in Alg.~\ref{alg:ODFM-CSVD-HFB} provides increasingly better value for the objective function in \eqref{eq:OFDM_analogBF}. Thus, the convergence is determined once the objective function ceases to increase. From our simulations, the algorithm require on average 4.36 iterations to find each analog precoder and combiner vector pair, and has converged with less than 10 iterations in nearly 95\% of the trials.
	
\vspace{-0.2cm}
\section{Conclusion}
\vspace{-0.12cm}
\label{sec:conclusion}

	We have proposed a novel hybrid~beamforming~design~for mmWave OFDM massive MIMO systems based on the constrained Tucker2 decomposition.
	The proposed constrained Tucker2 decomposition is used to design the analog precoder and combiner matrices by maximizing the sum of the effective baseband channel gain of each data stream in every subcarrier while reducing the interference among data streams within the same subcarrier. The digital precoder and combiner are obtained from the effective baseband channel's SVD on a per-subcarrier basis.
	Simulation results confirmed the effectiveness of the proposed design, which achieves near-optimal sum-rate over a wide range of SNR, number of data streams, and number of antennas, and outperforms other existing designs.


%

%
%
%
%

\ifCLASSOPTIONcaptionsoff
  \newpage
\fi


\vspace{-0.17cm}
\begin{thebibliography}{1}
\vspace{-0.16cm}


	\bibitem{Xiao2017}
	M.~Xiao \hspace{-0.1cm} \textit{et al.},\hspace{-0.1cm} ``Millimeter wave communications for future mobile net\-works,'' \textit{IEEE J. S. A. Commun.}, vol. 35, no. 9, pp. 1909--1935, \mbox{Sep. 2017}.
	
	\bibitem{Giordani2020}
	M.~Giordani \textit{et al.}, ``Toward 6G networks: use cases and technologies,'' \textit{IEEE Commun. Mag.}, vol. 58, no. 3, pp. 55--61, Mar. 2020.
		
	\bibitem{Pi2011}
	Z.~Pi and F.~Khan, ``An introduction to millimeter-wave mobile broadband systems,'' \textit{IEEE Commun. Mag.}, vol. 49, no. 6, pp. 101--107, Jun. 2011. 
	
	\bibitem{Rappaport2013}
	T.~S.~Rappaport \textit{et al.}, ``Millimeter wave mobile communications for 5G cellular: it will work!'' \textit{IEEE Access}, vol. 1, pp. 335--349, May 2013. 
		
	\bibitem{Heath2016}
	R.~W.~Heath \textit{et al.}, ``An overview of signal processing techniques for millimeter wave MIMO systems,'' \textit{IEEE J. Sel. Topics Signal Process.}, vol. 10, no. 3, pp. 436--453, Apr. 2016.
	
	\bibitem{Ayach2014}
	O.~El~Ayach \textit{et al.}, ``Spatially sparse precoding in millimeter wave MIMO systems,'' \textit{IEEE Trans. Wirel. Commun.}, vol. 13, no. 3, pp. 1499--1513, Mar. 2014.
	
	\bibitem{Sohrabi2016}
	F. Sohrabi, and W. Yu, ``Hybrid digital and analog beamforming design for large-scale antenna arrays,'' \textit{IEEE J. Sel. Top. Signal Process.}, vol. 10, no. 3, pp. 501--513, Apr. 2016.
	
	\bibitem{Zilli2020}
	G.~M.~Zilli and W.-P. Zhu, ``Constrained-SVD based hybrid beamforming design for millimeter wave communications,'' in IEEE 92nd Veh. Techn. Conf., Oct 2020, pp. 1--5.
	
	\bibitem{Bolckei2002}
	H.~Bolckei, D.~Gesbert, and A.~J.~Paulraj, ``On the capacity of OFDM-based spatial multiplexing systems,'' \textit{IEEE Trans. Commun.}, vol. 50, no. 2, pp. 225--234, Feb. 2002.
	
	\bibitem{Yu2016}
	\hspace{-0.2cm} X.~Yu, J.-C.~Shen, J.~Zhang, and K.~B.~Letaief, ``Alternating minimization algorithms for hybrid precoding in millimeter wave MIMO systems,'' \textit{IEEE J. Sel. Topics Signal Process}, vol. \hspace{-0.1cm} 10, \hspace{-0.1cm}  no. \hspace{-0.1cm} 3, \hspace{-0.1cm}  pp. \hspace{-0.1cm} 485--500, \hspace{-0.1cm} Apr. \hspace{-0.1cm}2016.
		
	\bibitem{Sohrabi2017}
	\hspace{-0.25cm} F. Sohrabi, and W. Yu, ``Hybrid analog and digital beamforming for mmWave OFDM large-scale antenna arrays,'' \textit{IEEE J. Sel. Areas Commun.}, vol. 35, no. 7, pp. 1432--1443, Jul. 2017.
	
	\bibitem{Tsai2019}
	\hspace{-0.2cm} T. Tsai, M. Chiu, and C. Chao, ``Sub-system SVD hybrid beamforming design for millimeter wave multi-carrier systems,'' \textit{IEEE Trans. Wirel. Commun.}, vol. 18, no. 1, pp. 518--531, Jan. 2019.
	
	\bibitem{Chiang2018}
	\hspace{-0.2cm} H.-L.~Chiang \textit{et al.}, ``Hybrid beamforming based on implicit channel state information for millimeter wave links,'' \textit{IEEE J. Sel. Topics Signal Process}, vol. 12, no. 2, pp. 326--339, May 2018.
	
	\bibitem{Araujo2019}	  
	\hspace{-0.2cm} D.~C.~Araujo \textit{et al.}, ``Tensor-based channel estimation for massive MIMO-OFDM systems,'' \textit{IEEE Access}, vol. 7, pp. 42133--42147, Mar. 2019.
	
	\bibitem{Ruble2020}
	\hspace{-0.2cm} M.~Ruble and I.~Guvenc, ``Multilinear singular value decomposition for millimeter wave channel parameter estimation,'' \textit{IEEE Access}, vol. 8, pp. 75592--75606, Apr. 2020.
	
	\bibitem{Almeida2008}
	\hspace{-0.2cm} A.~L.~F.~de~Almeida, G.~Favier, and J.~C.~M.~Mota, ``A constrained factor decomposition with application to MIMO antenna systems,'' \textit{IEEE Trans. Signal Process.}, vol. 56, no. 6, pp. 2429--2442, Jun. 2008.
	
	\bibitem{Liu2017}
	\hspace{-0.2cm} L.~Liu, and Y.~Tian, ``Hybrid precoding based on tensor decomposition for mmWave 3D-MIMO systems,'' in Proc. IEEE/CIC Int. Conf. Commun. China (ICCC), Qingdao, China, Oct. 2017, pp. 1--6.
		
	\bibitem{Lee2014}	
	\hspace{-0.2cm} J.~Lee~and~Y.~H.~Lee, ``AF relaying for millimeter wave communication systems with hybrid RF/baseband MIMO processing,'' in Proc. IEEE Int. Conf. Commun., Sydney, Australia, Jun. 2014, pp. 5838--5842.

	\bibitem{Kolda2009}
	\hspace{-0.2cm} T.~G.~Kolda and B.~W.~Bader, ``Tensor decompositions and applications,'' \textit{SIAM Rev.}, vol. 51, no. 3, pp. 455--500, Aug. 2009.

	\bibitem{Lathauwer2000b}
	\hspace{-0.2cm} L.~De~Lathauwer, B.~De~Moor, and J.~Vandewalle, ``On the best rank-1 and rank-($R_1$,$R_2$, $\cdots$, $R_N$) approximation of higher-order tensors'' \textit{SIAM J. Matrix Anal. Appl.}, vol. 21, no. 4, pp. 1324--1342, 2000.
	
	\bibitem{Lathauwer2000}
	\hspace{-0.2cm} L.~De~Lathauwer, B.~De~Moor, and J.~Vandewalle, ``A multilinear singular value decomposition,'' \textit{SIAM J. Matrix Anal. Appl.}, vol.~21, no.~4, pp. 1253--1278, 2000.
		
	
\end{thebibliography}
\end{document}